\documentclass[twocolumn,fleqn]{aastex631}
\usepackage{graphicx}	% Including figure files
\usepackage{mathrsfs}
\usepackage{amsmath}	% Advanced maths commands
\usepackage{bm}
\usepackage{tikz}
\usepackage{amssymb}	% Extra maths symbols
\usepackage{enumitem}   %FJ: for itemization without bullets and indents
\usepackage{soul} % FJ, for strike through text
\usepackage[hang]{footmisc}
\newcommand{\SatGen}{{\tt SatGen}\,}
\newcommand{\hmf}{{\tt hmf}\,}
\newcommand{\VVV}{{\tt VVV-L1}\,}
\newcommand{\GUREFT}{{\tt GUREFT}\,}

\newcommand{\eq}[1]{eq.~(\ref{eq:#1})}
\newcommand{\eqs}[1]{eqs.~(\ref{eq:#1})}

\newcommand{\se}[1]{Section \ref{sec:#1}}

\newcommand{\Fig}[1]{Fig.~\ref{fig:#1}}

\newcommand{\be}{\begin{equation}}
\newcommand{\ee}{\end{equation}}
\newcommand{\bad}{\begin{equation} \begin{aligned}}
\newcommand{\ead}{\end{aligned} \end{equation}}

\newcommand{\Msun}{M_\odot}

\newcommand{\Gyr}{\,{\rm Gyr}}
\newcommand{\cm}{\,{\rm cm}}
\newcommand{\g}{\,{\rm g}}
\newcommand{\kms}{\,{\rm km/s}}

\newcommand{\rhoc}{\rho_{\rm crit}}

\newcommand{\rhos}{\rho_{\rm s}}

\newcommand{\deltac}{\delta_{\rm c}}
\newcommand{\Mv}{M_{\rm vir}}
\newcommand{\Mbh}{M_{\rm bh}}
\newcommand{\dMbhdt}{\dot{M}_{\rm bh}}

\newcommand{\Mres}{M_{\rm res}}
\newcommand{\mv}{m_{\rm vir}}

\newcommand{\rv}{r_{\rm vir}}

\newcommand{\rs}{r_{\rm s}}

\newcommand{\sigmam}{\sigma_m}
\newcommand{\sigmaom}{\sigma_{0m}}
\newcommand{\sigmaeff}{\sigma_{\rm eff}}
\newcommand{\sigmaeffm}{\sigma_{{\rm eff},m}}

\newcommand{\Vmax}{V_{\rm max}}
\newcommand{\vtilde}{\tilde{v}}

\newcommand{\vmax}{v_{\rm max}}
\newcommand{\veff}{v_{\rm eff}}

\newcommand{\xc}{x_{\rm circ}}

\newcommand{\tcoll}{t_{\rm cc}}
\newcommand{\thalf}{t_{1/2}}

\newcommand{\tmerge}{t_{\rm merge}}
\newcommand{\tcross}{t_{\rm cross}}
\newcommand{\tlkbkform}{t_{\rm form}}
\newcommand{\tE}{t_{\rm E}}
\newcommand{\lambdaE}{\lambda_{\rm E}}
\newcommand{\rmd}{{\rm d}}

\newcommand{\fEPS}{f_{\rm EPS}}

\begin{document}
\nolinenumbers

\title{Formation of the Little Red Dots from the Core-collapse of Self-interacting Dark Matter Halos}

\author[0000-0001-6115-0633]{Fangzhou Jiang}\thanks{Corresponding author: \href{mailto:fangzhou.jiang@pku.edu.cn}{fangzhou.jiang@pku.edu.cn}}\thanks{Boya Young Fellow}
\affiliation{Kavli Institute for Astronomy and Astrophysics, Peking University, Beijing,  100871, China}

\author{Zixiang Jia}
\affiliation{Department of Astronomy, School of Physics, Peking University, 5 Yiheyuan Road, Beijing, 100871, China}

\author[0000-0002-1665-5138]{Haonan Zheng}
\affiliation{Kavli Institute for Astronomy and Astrophysics, Peking University, Beijing,  100871, China}

\author[0000-0001-6947-5846]{Luis C. Ho}
\affiliation{Kavli Institute for Astronomy and Astrophysics, Peking University, Beijing 100871, China}
\affiliation{Department of Astronomy, School of Physics, Peking University, 5 Yiheyuan Road, Beijing, 100871, China}

\author[0000-0001-9840-4959]{Kohei Inayoshi}
\affiliation{Kavli Institute for Astronomy and Astrophysics, Peking University, Beijing, 100871, China}

\author[0000-0002-6196-823X]{Xuejian Shen}
\affiliation{Department of Physics, Massachusetts Institute of Technology, Cambridge, 02139, MA, USA}
\affiliation{Kavli Institute for Astrophysics and Space Research, Massachusetts Institute of Technology, Cambridge, 02139, MA, USA}

\author[0000-0001-8593-7692]{Mark Vogelsberger}
\affiliation{Kavli Institute for Astrophysics and Space Research, Massachusetts Institute of Technology, Cambridge, 02139, MA, USA}

\author[0000-0002-9048-2992]{Wei-Xiang Feng}
\affiliation{Department of Physics, Tsinghua University, Beijing 100084, China}



\begin{abstract}
We present a statistical study of black hole (BH) formation and growth seeded by gravothermal core collapse of self-interacting dark matter (SIDM) halos at high redshift, using a cosmological semi-analytical framework based on Monte Carlo merger trees. 
We demonstrate that gravothermal collapse naturally leads to BH formation in high-concentration halos at a characteristic mass scale set by the SIDM cross section, and occurs predominantly in the early Universe.
This mechanism is particularly promising for explaining the abundance of the little red dots (LRDs) — a population of early, apparently galaxy-less active galactic nuclei hosting supermassive BHs. 
By incorporating this seeding process with simple models of BH growth and assuming a 100\% duty cycle, we reproduce the observed LRD mass function for velocity-dependent cross sections of \(\sigma_{0m} \sim 30\,\mathrm{cm^2\,g^{-1}}\) and \(\omega \sim 80\,\mathrm{km\,s^{-1}}\), which are consistent with independent constraints from local galaxies. 
While higher values of $\sigma_{0m}$ (or $\omega$) would overpredict the low-mass (or high-mass) end of the BH mass function, such deviations could be reconciled by invoking a reduced duty cycle or lower Eddington ratio. 
Our results suggest that the demographics of high-redshift BHs can serve as a novel and complementary probe of SIDM physics. 
\end{abstract}



\section{Introduction} \label{sec:intro}

Self-interacting dark matter (SIDM) has emerged as an appealing candidate to address longstanding cosmological challenges \citep{Weinberg15,Sales22}. 
Unlike the standard cold dark matter (CDM) model, SIDM introduces elastic scattering between dark matter (DM) particles, preserving CDM's successes on large scales \citep[e.g.,][]{Rocha13,Despali25} while significantly altering halo structures on smaller scales. 
These alterations manifest primarily in two phases driven by gravothermal evolution: first, the formation of an isothermal core that can potentially resolve the cusp-core discrepancy observed in dwarf galaxy rotation curves without invoking baryonic physics \citep{SpergelSteinhardt00,Kaplinghat14,Kaplinghat16}; second, a gravothermal instability leading to core-collapse and the eventual formation of central black holes (BHs) \citep{BalbergShapiro02, Pollack15}.

Recent theoretical advancements highlight SIDM core-collapse as a promising mechanism for seeding supermassive black holes (SMBHs) in the early Universe. 
\citet{Feng21, Feng22, Feng25} and \citet{GadNasr24} show via relativistic calculations that about 0.1-1\% of halo mass will turn into a BH seed, independent of the mass or structure of the halo at formation.  
\citet{Roberts25} indicates that rare, massive SIDM halos at high redshifts ($z \sim 6-10$) can rapidly collapse to form seed BHs as massive as $\sim10^7$–$10^9\Msun$, potentially explaining the existence of extremely massive quasars at cosmic dawn. 
However, beyond these extreme systems, a more pervasive observational challenge has emerged with the discovery of little red dots (LRDs) — a new population of compact, high-redshift ($z\gtrsim5$) active galactic nuclei that appear unusually abundant \citep{Kokorev24,Kocevski24,Maiolino24} yet remarkably deficient in stellar components \citep{Chen24}. 
These enigmatic systems, likely hosting SMBHs with masses ranging from $10^6$ to $10^8\Msun$ \citep{Matthee24,Greene24,Taylor24}, pose significant challenges to traditional baryonic seeding scenarios \citep{Inayoshi20,Volonteri21} and conventional galaxy-BH co-evolution pictures \citep{KormendyHo13, ReinesVolonteri15}.

It is thus interesting to explore the possibility that LRDs originate from DM via SIDM core-collapse. 
If core-collapse occurs prior to re-ionization, BHs can form in halos that are devoid of significant stellar components. 
The critical question then arises: what conditions must be met for this dark seeding mechanism to take place?
Qualitatively, core-collapse must occur early enough, while the DM halos responsible for seeding these BHs must be sufficiently massive to account for the observed BH masses. 
These two conditions are in competition due to the hierarchical nature of structure formation. 
Furthermore, the timing of core-collapse is intricately linked to the cross section of dark self-scattering, as the cross section, along with halo mass and structure at formation, determines the timescale for core-collapse \citep{BalbergShapiro02,Pollack15,Yang23}.

In this Letter, we present a proof-of-concept study addressing this question. 
In \se{CharacteristicMass}, we demonstrate the existence of a characteristic halo mass scale at which SIDM core-collapse occurs most rapidly.  
In \se{seeding}, we further clarify the conditions required for BH seeding, exploring which halos can collapse sufficiently early to explain the observed properties of LRDs.
We introduce a semi-analytic framework for modeling BH seeding and subsequent growth in \se{model}.
Using this model, we compute the predicted BH mass function in \se{MassFunction}, illustrating that SIDM scenarios can naturally reproduce the observed LRD populations.
Finally, in \se{discussion}, we discuss how the SIDM cross section accommodating the LRD population relates to completely independent constraints from galaxy kinematics, and we summarize our finding in \se{conclusion}.

Throughout this study, we define the virial radius of a DM halo as the radius enclosing an average density equal to 200 times the critical density of the Universe.
We adopt a flat cosmological model characterized by the present-day matter density $\Omega_m=0.3$, baryonic density $\Omega_b=0.0465$, dark-energy density $\Omega_\Lambda=0.7$, power-spectrum normalization $\sigma_8=0.8$, spectral index $n_s=1$, and a Hubble parameter $h=0.7$. 
Following standard convention, we express the DM self-interaction cross section in terms of the cross section per unit particle mass, $\sigmam = \sigma/m_\chi$, where $m_\chi$ is the mass of a DM particle.
We consider velocity-dependent cross sections specified by two parameters, a low-velocity cross section $\sigmaom$ and a characteristic velocity scale $\omega$, above which cross section rapidly declines \citep{YangYu22}. 

%-----------------------------------------------------------------

\section{Characteristic halo mass for fastest core-collapse} \label{sec:CharacteristicMass}

\begin{figure}
    \centering
        \includegraphics[width=\linewidth]{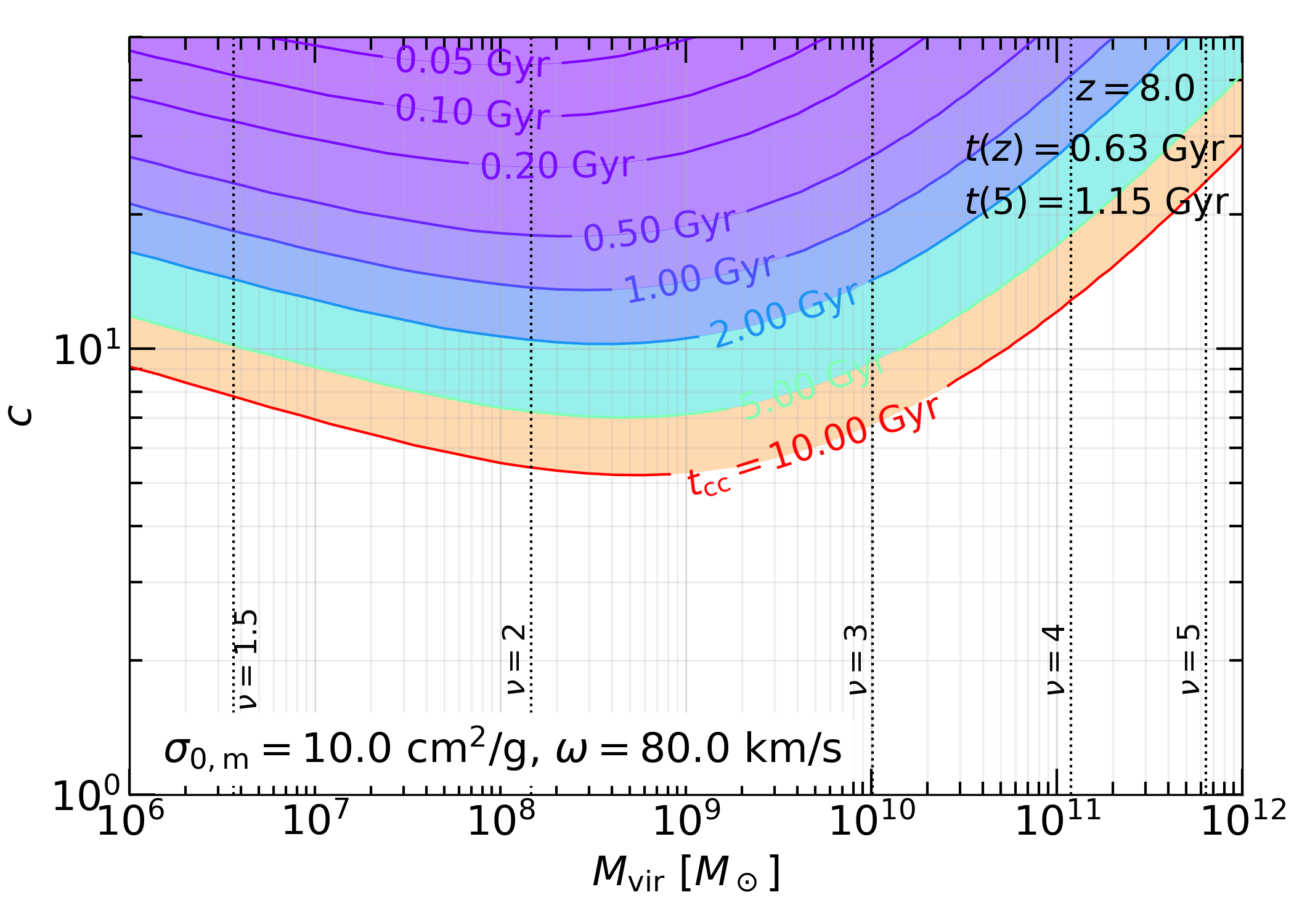}
        \includegraphics[width=\linewidth]{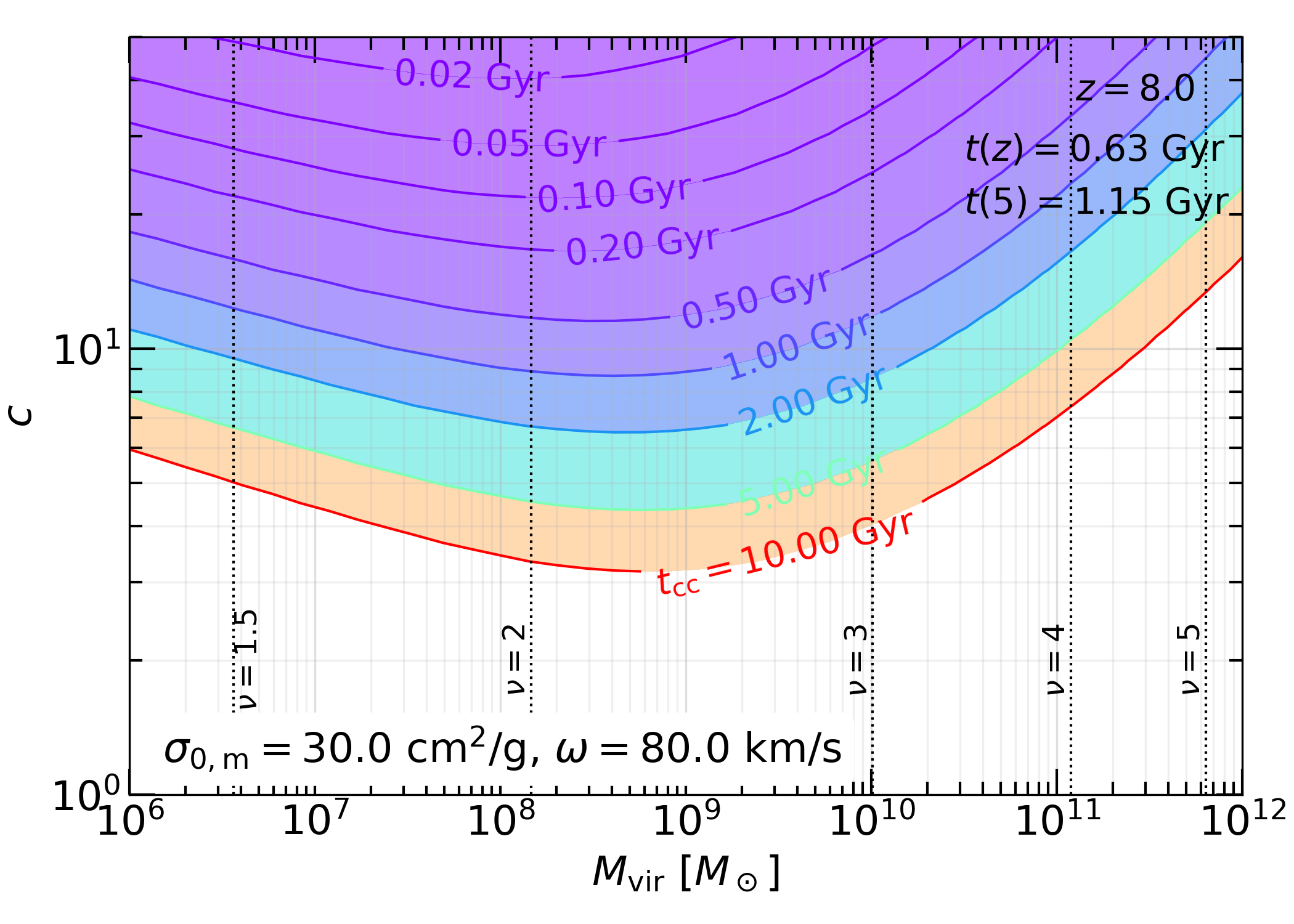}
        \includegraphics[width=\linewidth]{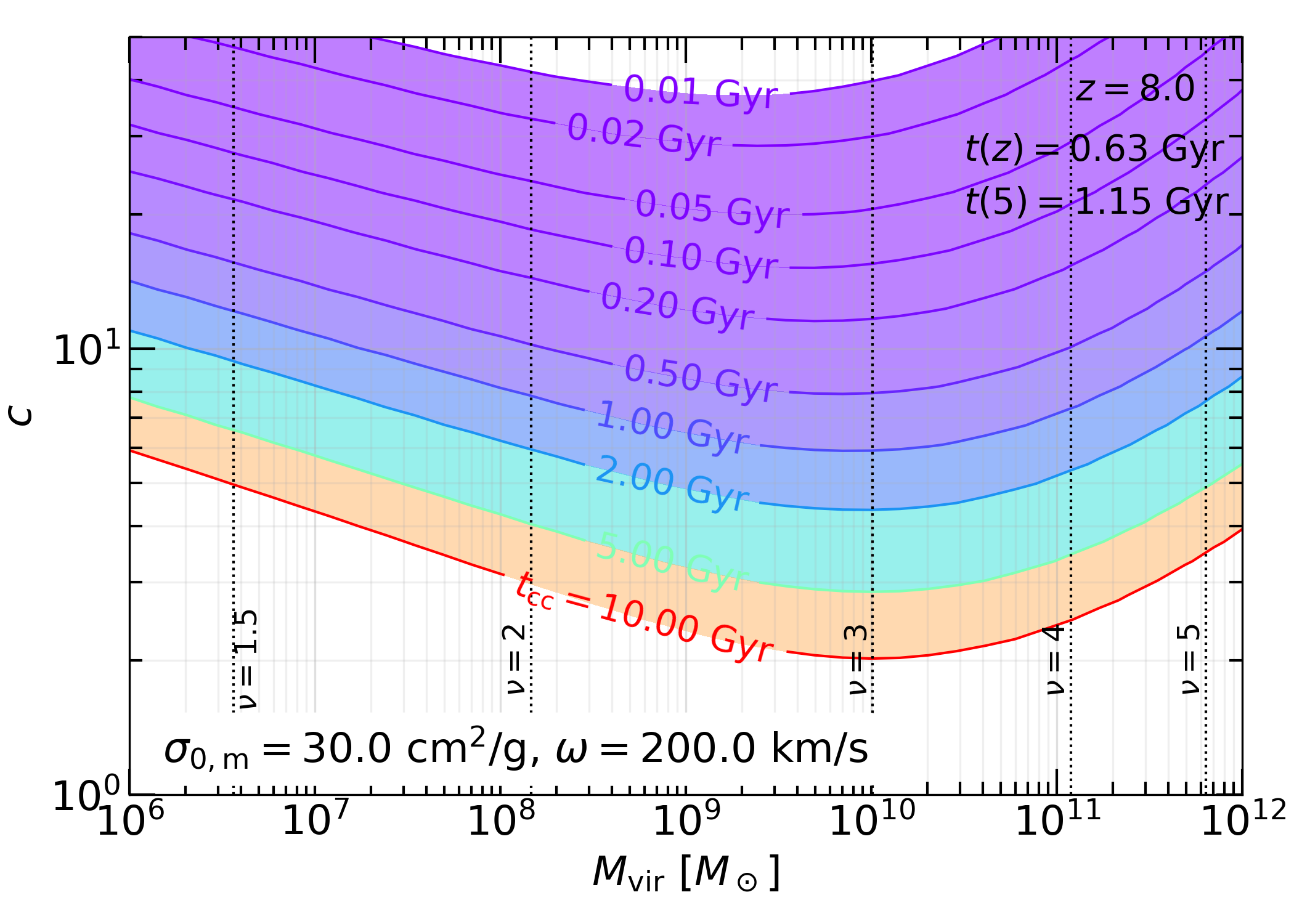}
    \caption{
    Contours of SIDM core-collapse timescale  $\tcoll$ as a function of halo concentration $c$ and halo mass $\Mv$ at redshift $z=8$. 
    Cross sections are indicated in the lower-left corners. 
    Cosmic times at $z=8$ and $z=5$ are labeled in the upper-right corners.
    Halos located above a given contour, $\tcoll \la (1.15 - 0.63)\Gyr \approx 0.5\Gyr$, can undergo core-collapse by the time of $z=5$, representative of LRDs. 
    Clearly, timely core-collapse requires halos with high concentrations ($c \sim 10$). 
    Vertical dotted lines indicate halo rarity in terms of their density peak height  $\nu$. 
    The characteristic halo mass corresponding to the fastest core-collapse aligns with rare 2-3$\sigma$ peaks, shifting to higher masses with increasing effective cross section (top to bottom panels).
    }
    \label{fig:ConcentrationMass}
\end{figure}

To seed a BH from an SIDM halo at high redshift, the characteristic timescale for gravothermal core-collapse must be very short.
Since this timescale is sensitive to the self-interaction cross section, we first clarify our definition of the DM scattering cross section before elaborating on the core-collapse time.

Consider an SIDM particle $\chi$ coupled to a light gauge boson $\phi$. 
In the perturbative (Born) regime, the differential cross section for elastic self-scattering in the center-of-momentum frame, assuming Rutherford-type scattering, is given by \citep{YangYu22}:
\be\label{eq:CrossSection}
\frac{\rmd \sigma}{\rmd \cos \theta} = \frac{\sigma_0\omega^4}{2[\omega^2+v^2\sin^2(\theta/2)]^2}
\ee
where $\sigma_0=4\pi\alpha_\chi^2/( m_\chi^2\omega^4)$ is the low-velocity limit of the cross section, $\alpha_\chi$ is the coupling strength, $\omega=m_\phi c/m_\chi$ is the velocity scale above which cross section sharply declines, $\theta$ is the scattering angle, and $v$ is the relative velocity between two DM particles. 

For an SIDM halo, it is useful to define an effective, velocity-averaged cross section by integrating the velocity-dependent cross section over the velocity distribution \citep{YangYu22,Yang23}:
\be\label{eq:EffectiveCrossSection}
\sigmaeff =\frac{1}{2}\int \vtilde^2 \rmd \vtilde \sin^2\theta\rmd\cos\theta \frac{\rmd \sigma}{\rmd \cos\theta} \vtilde^5 e^{-\vtilde^2},
\ee
where $\vtilde\equiv v/(2\veff)$, and $\veff$ is a characteristic one-dimensional velocity dispersion.
For an isotropic velocity field, it can be approximated by $\veff\approx \vmax/\sqrt{3}$, with $\vmax$ being the maximum circular velocity of the halo. 

Following previous studies \citep{BalbergShapiro02,Pollack15,Essig19,Yang23}, we approximate the core-collapse time as:
\be\label{eq:CoreCollapseTime}
\tcoll \approx \frac{150}{C}\frac{1}{\sigmaeffm\rhos\rs}\frac{1}{\sqrt{4\pi G\rhos}},
\ee
where $C$ is an empirical constant calibrated with $N$-body simulations, for which we adopt $C=0.75$, consistent with \citet{Nishikawa20}\footnote{This fudge factor, of order unity, is related to the thermal conductivity of the gravothermal fluid and may depend on the detailed structure of the halo.}, and $\sigmaeffm$ is the effective cross section per particle mass.
This expression implies that the lifetime of the halo core prior to collapse is a multiple of the collisional relaxation time, which scales as $(G\rhos)^{-1/2}$, evaluated at the time of halo formation. 
We assume the halo follows a Navarro, Frenk \& White (NFW) profile \citep{NFW97} at the time of formation: 
\be
\rho(r) = \frac{\rhos}{r/\rs(1+r/\rs)^2},
\ee
where $\rhos$ and $\rs$ are the scale density and scale radius, respectively.
An NFW halo can be fully specified by these two parameters, or equivalently by its virial mass  $\Mv$ and concentration parameter $c\equiv \rv/\rs$, where $\rv$ is the virial radius. 
These are related through:
\be
\Mv = 4\pi \rhos \rs^3 f(c)
\ee 
and 
\be \label{eq:ScaleDensity}
\rhos = \frac{c^3}{3f(c)} \Delta\rhoc(z),
\ee 
with $f(x) = \ln(1+x) - x/(1+x)$, $\Delta=200$, and $\rhoc(z)$ the critical density of the Universe at redshift $z$.
To leverage established theoretical models describing the evolution of halo virial mass (e.g., \citealt{LC93}), we use $\Mv$ and $c$ to define the NFW profile of the halo at its formation epoch.

From \eqs{CoreCollapseTime}-(\ref{eq:ScaleDensity}), it is clear that the core-collapse timescale decreases with increasing halo concentration and self-interaction cross section. 
Less obvious, however, is a non-monotonic dependence on halo mass. 
\Fig{ConcentrationMass} illustrates these trends, showing contours of  $\tcoll$ in the parameter space defined by halo concentration $c$ and virial mass  $\Mv$.
For illustrative purposes, we consider a scenario where halo virialization occurs at $z\sim8$. 
We also consider a reference redshift of $z=5$, typical for LRDs, so the time interval is $\sim0.5\Gyr$.
Several insights into BH seeding conditions emerge from this figure. 

First, a very high halo concentration — at least  $\gtrsim10$ — is required for the core-collapse time to fall below $\sim0.5\Gyr$. 
If gravothermal evolution begins at $z=8$ such a collapse timescale would bring the system to $z \sim 5$, just early enough for the resulting BH to be observed as an LRD. 
While the average halo concentration at these redshifts is typically $\sim3$ \citep[e.g.,][]{Yung24}, concentrations as high as  $c\sim10$ is still attainable, as we will discuss shortly.

Second, for a fixed cross section and concentration, there exists a characteristic halo mass at which core-collapse is most efficient.
If the halo mass is significantly below this characteristic value, the central density is too low for effective self-scattering. Conversely, if the mass is too high, the cross section becomes too small due to the strong velocity dependence described in \eq{CrossSection}.
Interestingly, for the cross-section range explored here ($\sigmaom\sim 10$-$30\cm^2\g^{-1}$, $\omega\sim80$-$200\kms$), this characteristic mass lies in the range
 $\Mv\sim10^{8-10}\Msun$ — assuming that roughly $1\%$ of the halo mass ends up in the BH \citep{Feng21}, this yields seed BH masses of $\Mbh\sim10^{6-8}\Msun$, already comparable to those powering LRDs.  
However, halos with this characteristic mass are relatively rare at cosmic dawn. 
Specifically, a virial mass of $\Mv\sim10^{8-10}\Msun$ corresponds to a peak height of $\nu\sim2-3$, where $\nu\equiv \deltac(z)/\sqrt{S(\Mv)}$, with $\deltac(z)$ denoting the critical linear overdensity for halo formation and $S(\Mv)$ the variance of smoothed density field on the mass scale $\Mv$.
As we will show shortly, the halos that can undergo core-collapse {\it in time} to seed BHs are typically of lower mass.

Besides mass and concentration, could other halo structural parameters influence the core-collapse timescale? In principle, all halos acquire some angular momentum from large-scale tidal torques \citep{Bullock01}. 
However, the angular momentum in the central region of an SIDM halo dissipates efficiently due to self-interactions (i.e., effective dark viscosity), and has only a marginal impact on the core-collapse time \citep{Feng21}.

%-----------------------------------------------------------------

\section{The BH-seeding condition} \label{sec:seeding}

\begin{figure*}
    \centering
        \includegraphics[width=0.329\linewidth]{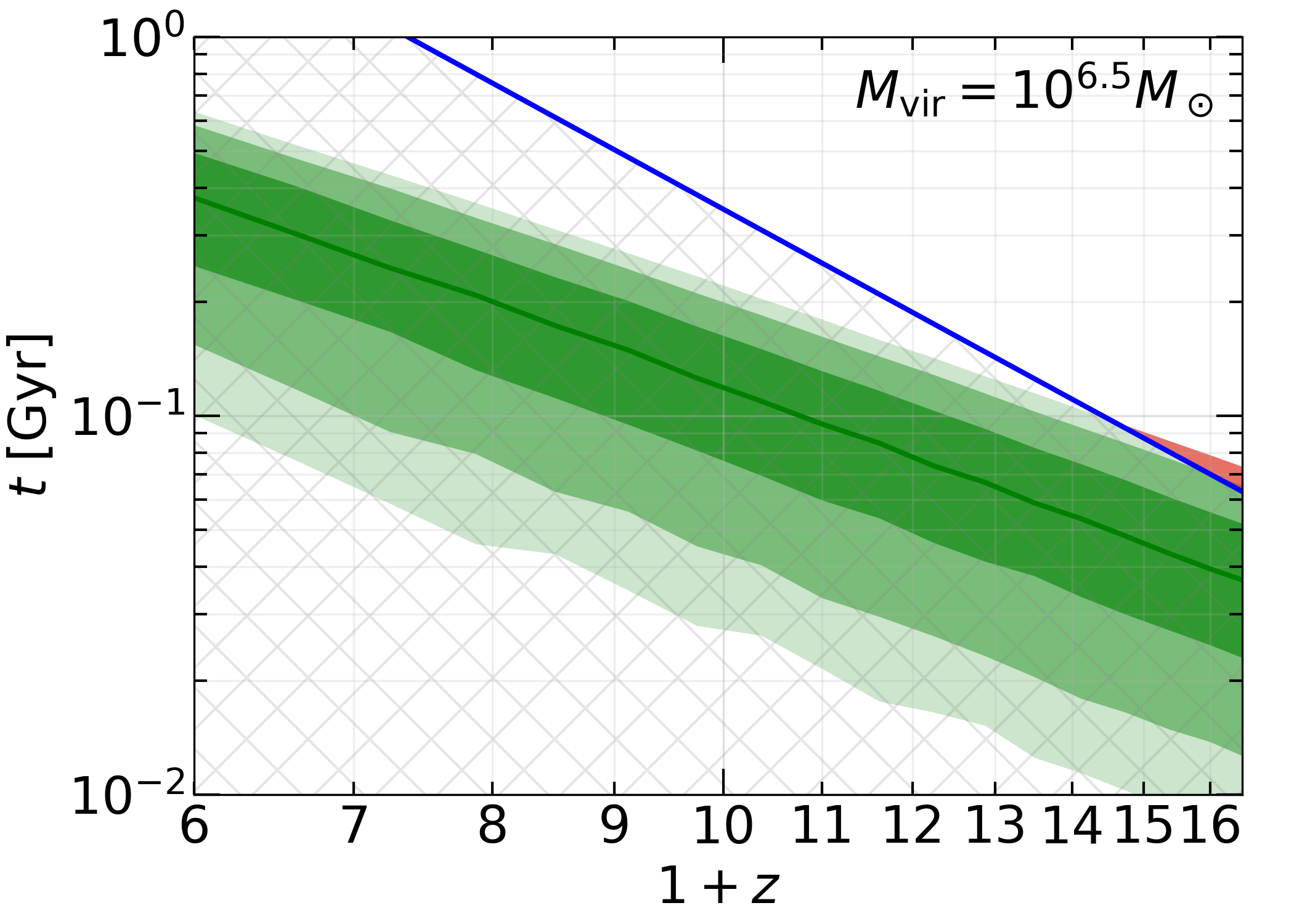}
        \includegraphics[width=0.329\linewidth]{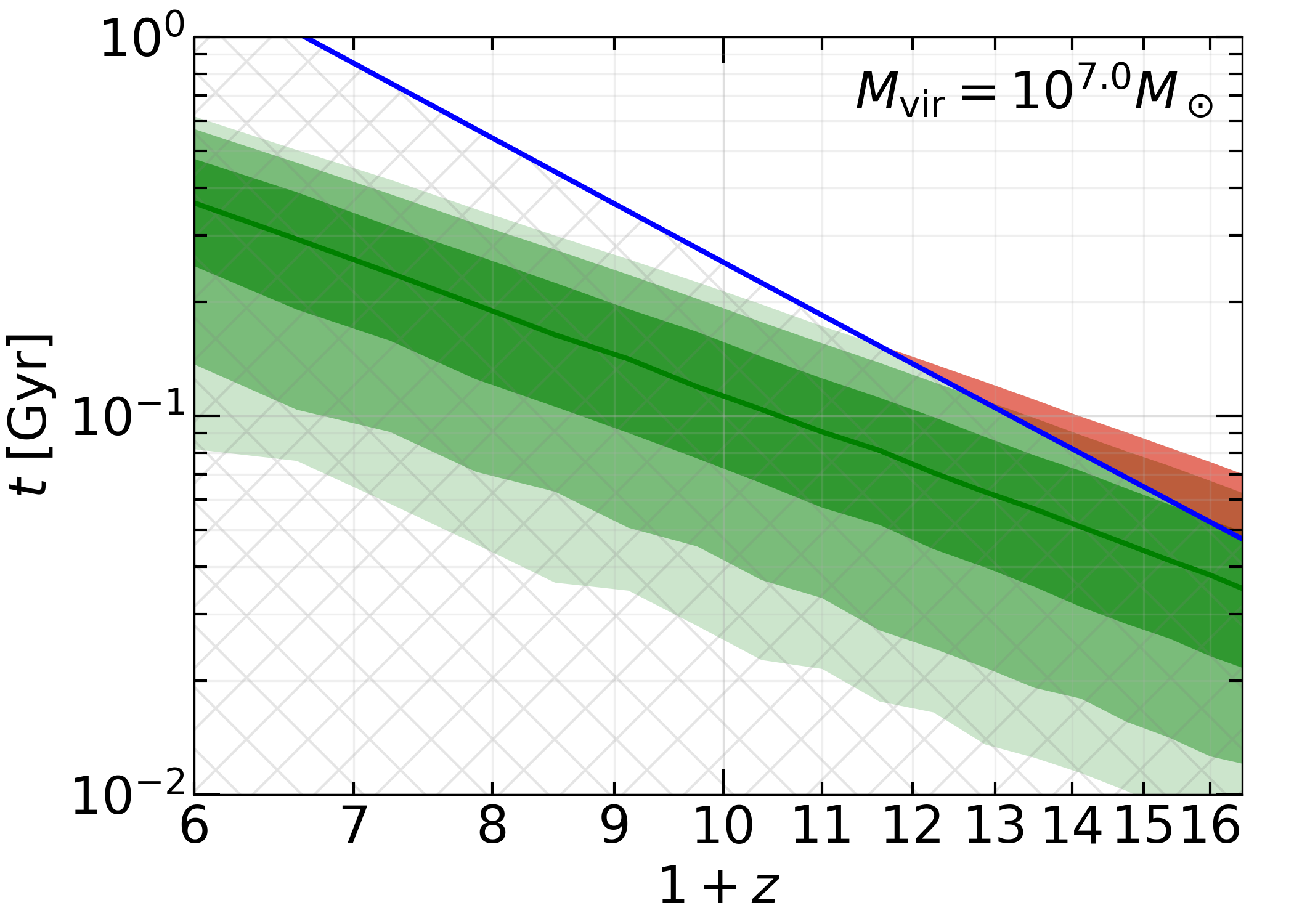}
        \includegraphics[width=0.329\linewidth]{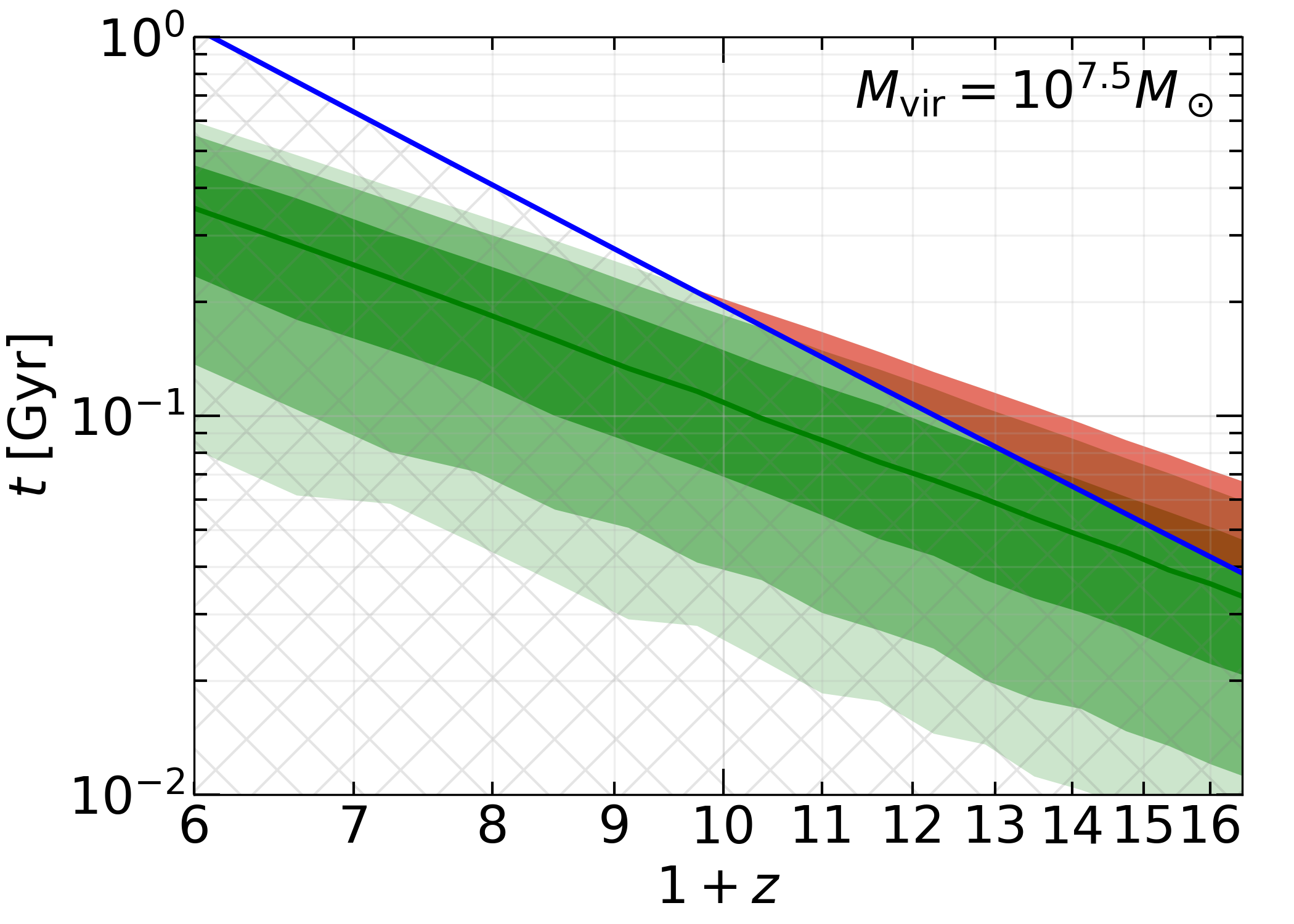}
        \includegraphics[width=0.329\linewidth]{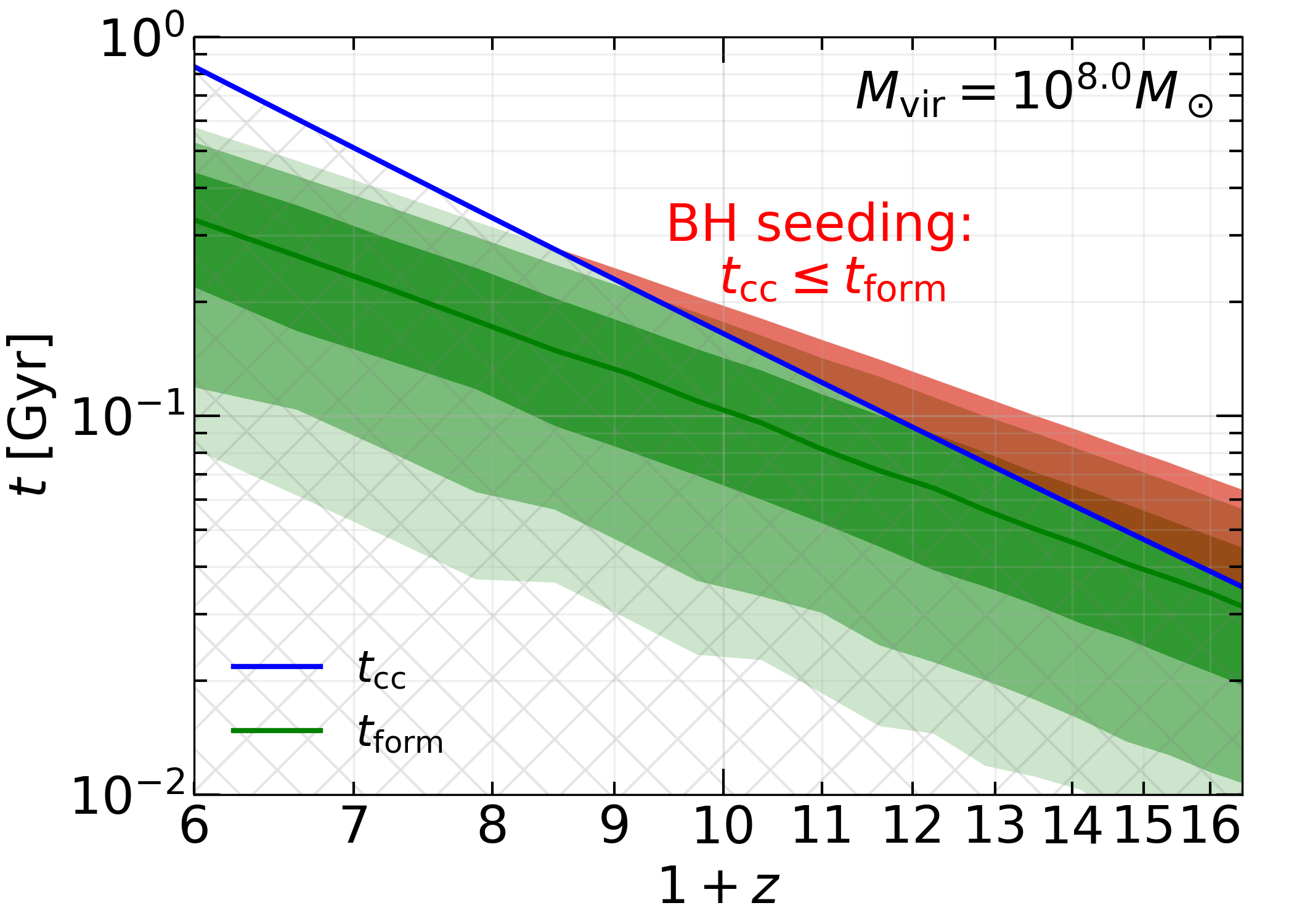}
        \includegraphics[width=0.329\linewidth]{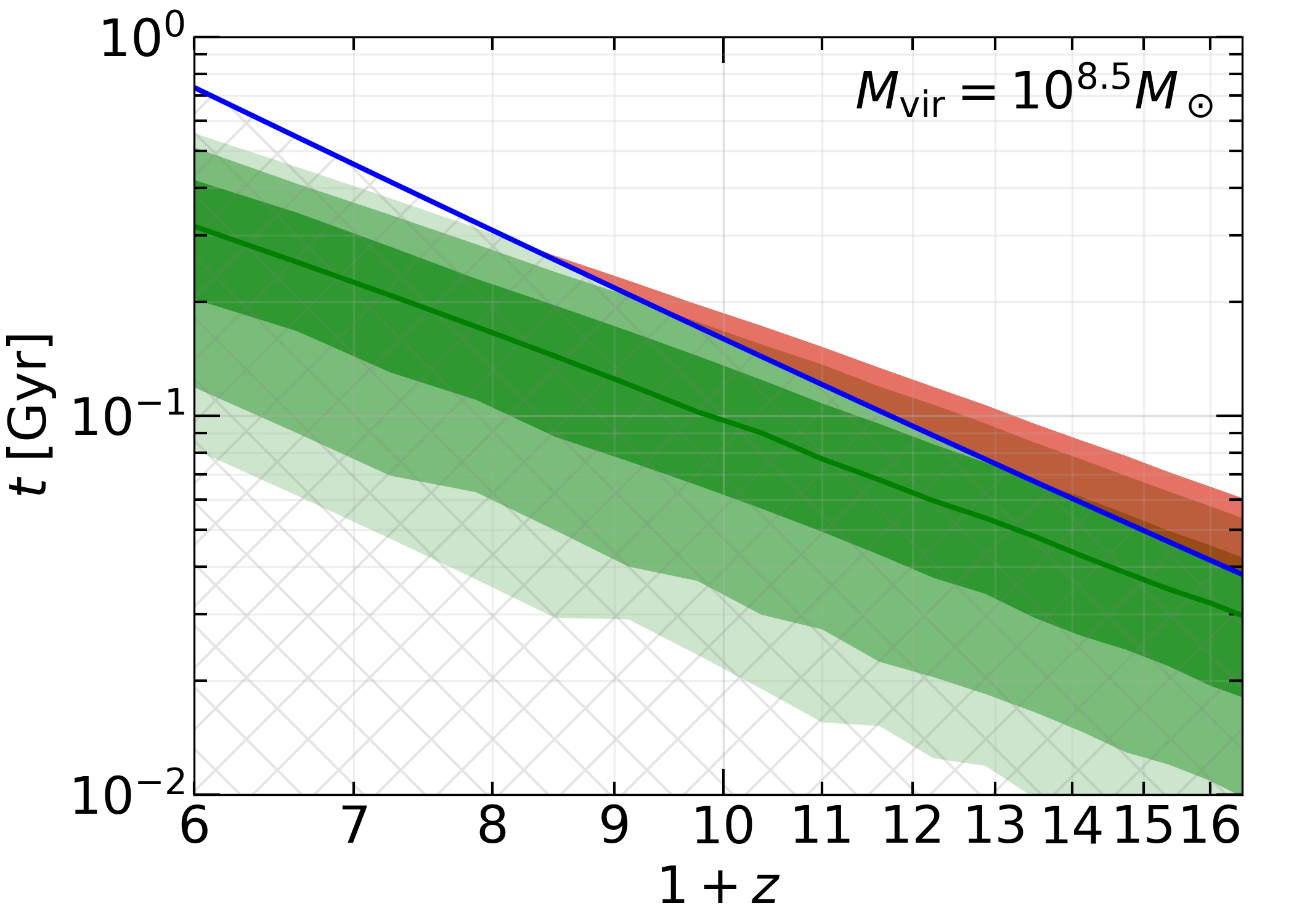}
        \includegraphics[width=0.329\linewidth]{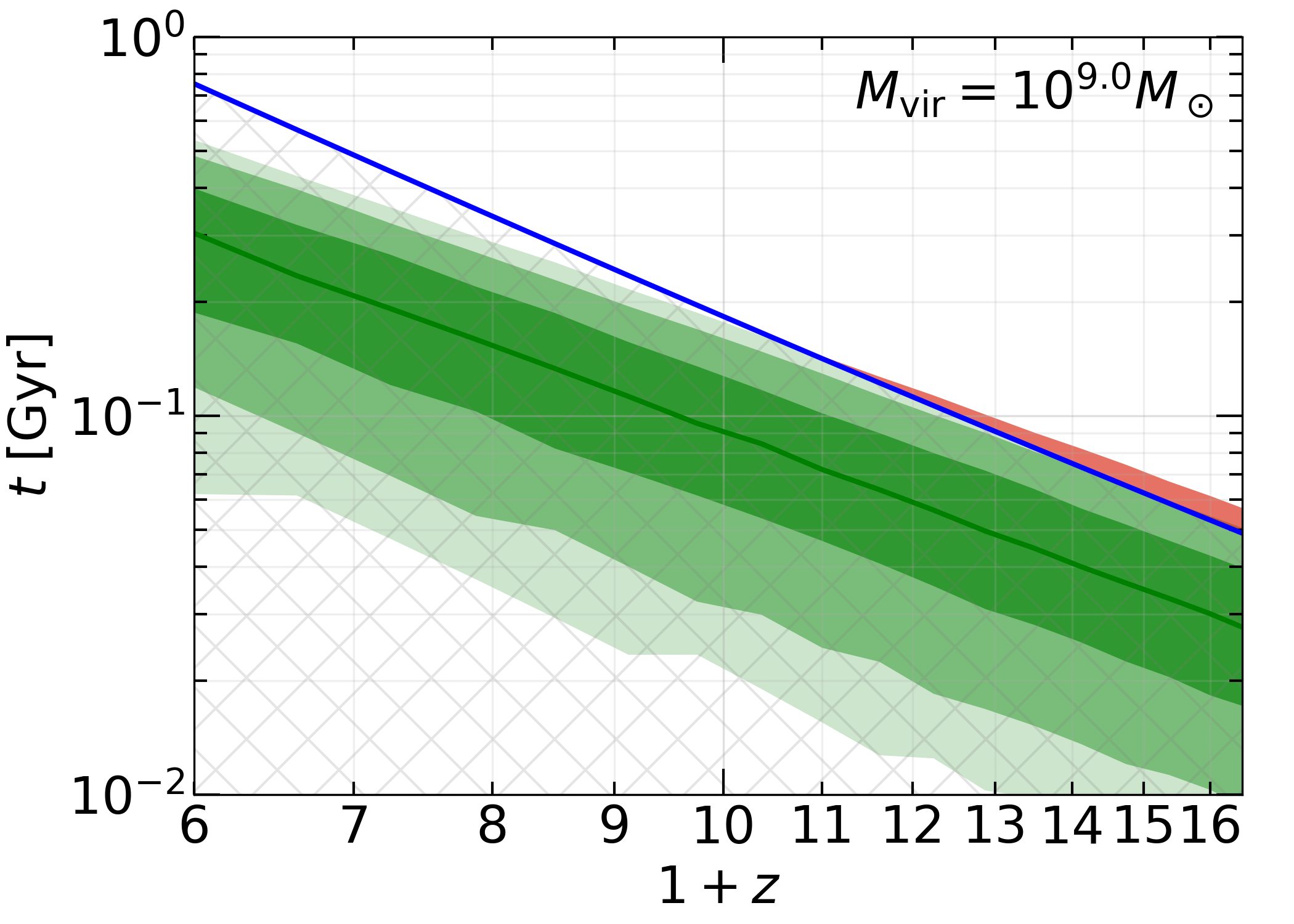}
    \caption{
    Exploration of the parameter space for BH seeding via core-collapse of SIDM halos, shown as a function of halo mass $\Mv$ and redshift $z$. 
    Each panel corresponds to a different instantaneous halo mass. 
    The halo concentration is fixed at $16$.
    The SIDM cross section is set to be $\sigmaom=30\cm^2/\g$ and $\omega=80\kms$. 
    Green bands indicate the distribution of look-back times to halo formation, $\tlkbkform(z,\Mv)$, predicted from the extended Press-Schechter formalism, with different shades representing the 25-75th, 10-90th, and 5-95th percentiles. 
    Blue lines show the core-collapse time $\tcoll(z)$, as defined in \eq{CoreCollapseTime}. 
    The hatched regions denote where the time since halo formation is insufficient for core-collapse to occur ($\tcoll>\tlkbkform$). 
    The early-forming tail, where collapse, and hence BH seeding, is possible ($\tcoll\leq\tlkbkform$), is highlighted in red. 
    This BH-seeding mechanism operates only at high redshift (in this case, $z\gtrsim8.5$; the exact threshold depends on concentration and cross section). 
    For seeding to occur within the redshift range explored here, the halo mass must lie in the range $\Mv\sim10^{7-8.5}\Msun$. 
    }
    \label{fig:BHseedingCondition}
\end{figure*}

To further clarify the conditions for BH formation via SIDM core-collapse, we compare two key timescales: the core-collapse time, $\tcoll$, as defined above, and the {\it look-back time to the epoch of halo formation}, $\tlkbkform$.
This comparison is motivated by the fact that gravothermal evolution driven by dark self-interactions only begins after a halo has formed.
An SIDM halo is expected to undergo core-collapse and seed a BH if $\tcoll<\tlkbkform$ — that is, if there is sufficient time since formation for the collapse to occur. 
We define the halo formation epoch as the cosmic time by which a halo has assembled half of its instantaneous mass\footnote{We note that this definition is somewhat arbitrary and could be refined through cosmological SIDM simulations that more accurately determine when gravothermal evolution begins. In practice, modelers might consider alternative markers such as the time of the last major merger, or when the halo first assembled the mass enclosed within its instantaneous scale radius $\rs$.}, denoted $\thalf$.  
The look-back time to formation is then 
\be\label{eq:FormationTime}
\tlkbkform(z,\Mv) = t(z) - \thalf(z,\Mv),
\ee
where $t(z)$ is the cosmic time at redshift $z$.

The halo-formation time depends on both halo mass and redshift, and can be computed analytically using the extended Press–Schechter (EPS) theory \citep[EPS, e.g.,][]{LC93}. 
For a halo of mass $M_0$ at cosmic time $t_0$, EPS predicts the ensemble-averaged number of progenitors with mass in the interval $[M_1, M_1+\rmd M_1]$ at an earlier time $t_1$:
\be\label{eq:PMF}
\frac{\rmd N}{\rmd M_1} \rmd M_1 = \frac{M_0}{M_1} \fEPS(S_1,\delta_1|S_0,\delta_0)  \left|\frac{\rmd S_1}{\rmd M_1}\right|\rmd M_1,
\ee
where 
\be\label{eq:FirstUpCrossingFraction}
\fEPS (S_1,\delta_1|S_0,\delta_0) =  \sqrt{\frac{1}{2\pi}}\frac{\Delta\delta}{(\Delta S)^{3/2}} \exp{\left[-\frac{(\Delta \delta)^2}{2\Delta S}\right]},
\ee
with $\delta = \deltac(z)$ denoting the linear critical overdensity for halo formation ($\Delta\delta=\delta_1-\delta_0$), and $S(M)$ the mass variance of the cosmic density field ($\Delta S=S_1-S_0$). 
Because a halo can have at most one progenitor in the mass range $M_1 \in [M_0/2, M_0]$,  the probability that a halo formed earlier than time $t_1$ (i.e., had already assembled at least half its mass by then) is given by:
\be\label{eq:FormationTimeCumulativeDistribution}
P(<t_1|M_0,t_0) = \int_{M_0/2}^{M_0} \frac{\rmd N}{\rmd M_1} \rmd M_1.
\ee
By substituting $M_0$ with halo mass $\Mv$, $t_0$ with cosmic time $t(z)$, and $t_1$ with the formation time $\thalf$, we can differentiate \eq{FormationTimeCumulativeDistribution} with respect to $\thalf$ to obtain the distribution of halo formation times.
This also yields the distribution of the look-back time to halo formation, $\tlkbkform(z,\Mv)$, which is shown in \Fig{BHseedingCondition} as green bands.

In \Fig{BHseedingCondition}, we compare the look-back time to halo formation, $\tlkbkform$,  with the core-collapse timescale, $\tcoll$, to identify the mass and redshift ranges favorable for BH formation.
For illustration, we fix halo concentration at $c=16$ and the SIDM parameters at $(\sigmaom,\omega)=(30\cm^2/\g, 80\kms)$, while scanning halo masses from $\Mv\sim10^{6.5}\Msun$ to $10^{9}\Msun$. 

BH formation is not possible if the time since halo formation is insufficient for core-collapse — that is, when $\tcoll>\tlkbkform$.
This region is hatched in \Fig{BHseedingCondition}, and it encompasses the majority of high-redshift halos.
Only a subset of early-forming halos, those in a mass-dependent tail where $\tcoll\leq\tlkbkform$, can undergo core-collapse in time to seed BHs.

While the specific mass and redshift thresholds depend on the chosen values of halo concentration and cross section, it is clear that this SIDM-based BH seeding mechanism only operates at high redshifts — in this case, $z\gtrsim8.5$.
This is both important and reassuring: if the mechanism remained active at later times, we would expect high occupation fractions of massive BHs in low-mass galaxies, in tension with low-redshift observations \citep{Ho08,KormendyHo13,Greene20}.

Another key insight from this analysis is that the BH-seeding mass is lower than the characteristic mass for the most rapid core-collapse. 
In this example, only halos with $\Mv\sim10^{6.5-8.5}\Msun$ are able to seed BHs in the redshift range explored, whereas the fastest core-collapse occurs at $\sim 10^{9}\Msun$. 
We note that our analytic estimate of the characteristic mass range for core-collapse is in excellent agreement with the semi-analytic estimate of \citet{Ando24} and is also consistent with what is revealed in the SIDM zoom-in simulations of \citet{Nadler25}, although these numerical studies focus on SIDM subhalos. 

Assuming a 1\% conversion efficiency from halo mass to BH mass \citep{GadNasr24}, this implies a seed BH mass range of $\Mbh\sim 10^{4.5-6.5}\Msun$.
This is an order of magnitude higher than what baryonic seeding mechanisms predict \citep{Li23,Li24}, but still lower than the BH mass estimates for LRDs, which spans $10^{6-8}\Msun$. 
Thus, while SIDM core-collapse can naturally produce relatively massive BH seeds, substantial growth may still be required to match the observed BH mass distribution. 
We notice though that the BH mass estimates are still highly uncertain \citep[e.g.,][]{Naidu25}, so the extent of further growth can be revisited once the observational estimates sharpen. 

%-----------------------------------------------------------------

\section{Semi-analytic model for BH seeding and growth}\label{sec:model}

We develop a model for BH seeding and mass growth by incorporating the aforementioned dark-seeding conditions into halo merger trees, along with prescriptions for BH accretion and mergers.

Our first step is to generate cosmological populations of DM halos at early times.
This is technically challenging because we focus on the high-concentration, early-forming tail of low-mass halos, with masses as small as  $\Mv\sim10^6\Msun$. 
Numerical simulations capable of capturing both the necessary resolution and large cosmic volume would be prohibitively expensive. 
Therefore, we adopt a semi-analytic approach based on Monte Carlo merger trees and analytical halo mass functions (HMFs).

Specifically, we construct merger trees using the algorithm of \citet{PCH08}, as implemented in the semi-analytic framework \SatGen \citep{Jiang21}, which has been shown to produce merger statistics consistent with cosmological simulations \citep{JB14}.
We select a final redshift of $z=5$ and generate $N_i$ merger trees for final halo masses $M_{{\rm vir},i}=10^{8,\, 8.25,\,8.5,\, ...,\, 11.5}\Msun$. 
The number of trees $N_i$ is set to 32 for the most massive final halos, and increases towards lower masses in proportion to the HMF at $z=5$, as computed using the {\tt hmf} calculator \citep{Murray13}.  
%This mass range is chosen to encompass the expected masses of LRD host halos \footnote{We do not include halos with lower final virial masses because the merger trees for halos with \( \Mv(z=5) > 10^8\,\Msun \) already span progenitor masses down to \( 10^6\,\Msun \) at higher redshifts, where core-collapse is possible.}, based on the observationally inferred BH masses of $10^{6-8}\Msun$ \citep{Matthee24, Greene24} and assuming $\Mbh \sim 0.01 \Mv$ \citep{Feng21}. 
For each progenitor halo in the merger trees, we trace the full formation history (i.e., all branches) down to a resolution mass of  $\Mres = 10^{6}\Msun$.
To ensure accurate tracking of halo formation times, we further extend the main branch near this resolution limit down to a {\it leaf} mass of $10^{4}\Msun$. 
The resulting merger trees provide the virial masses of all progenitor halos, $\Mv(z)$, at  $z > 5$, enabling us to evaluate when and where BH seeding can occur.

\begin{figure}	
\includegraphics[width=1.0\columnwidth]{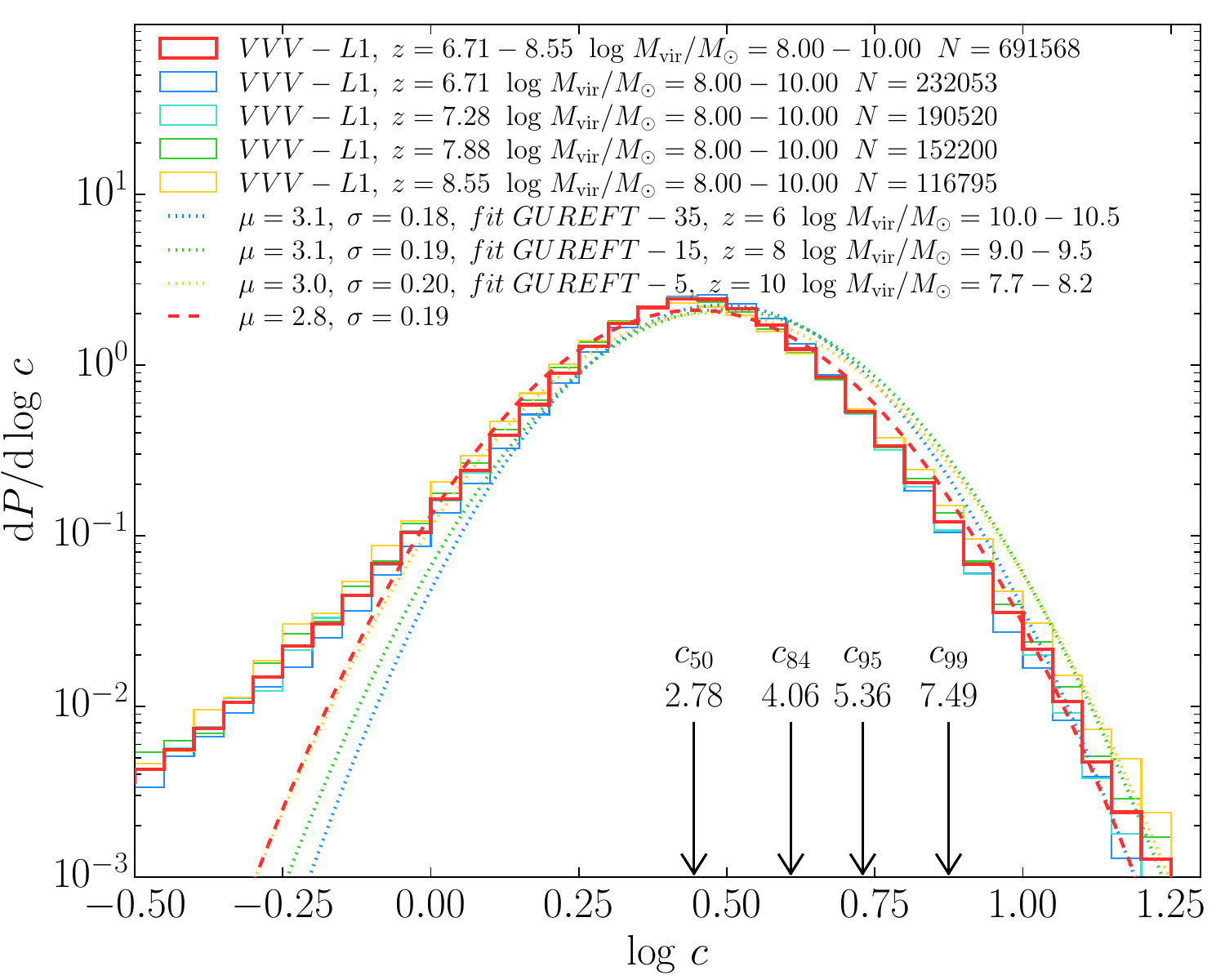}
    \caption{
    Distribution of halo concentration $c$ in the early Universe, measured from the \VVV cosmological simulation \citep{Wang20}. 
    No significant redshift evolution is observed across the redshift range, and the mass dependence is minimal over the simulation’s mass range of $10^{8-10}\Msun$.
    Arrows indicate the 50th, 84th, 95th, and 99th percentiles of the full distribution. 
    The high-$c$ tail is well described by a log-normal distribution with a median $c=2.8$ and a scatter $\sigma_{\log(c)}\approx 0.19$. 
    This distribution is used to assign concentrations to progenitor halos in our merger trees (see text).
    The presence of the high-$c$ tail is robust: the dotted lines show the best-fit log-normal distributions for the \GUREFT cosmological simulation \citep{Yung24}, which closely match the \VVV results.}
    \label{fig:ConcentrationDistribution}
\end{figure}

Second, we assign a concentration parameter $c$ to each progenitor branch. 
To do so, we measure the distribution of halo concentrations in the high-resolution cosmological simulation, \VVV \citep{Wang20}, applying the method of \citet{Wang24} to all well-resolved, relaxed  halos \footnote{We consider a halo {\it relaxed} if its substructure fraction is less than $0.1$ and the offset between its gravitational-potential minimum and its center-of-mass is less than $0.07 \rv$ \citep{Neto07}. We consider a halo {\it well-resolved} if it has more than 100 particles, and with the \citet{Wang24} method, we are able to achieve concentration convergence within 10\% for such halos.}.
We find that the distribution of $c$ shows negligible dependence on redshift at high $z$, as illustrated in \Fig{ConcentrationDistribution}, and almost no dependence on halo mass either (not shown here). 
The high-$c$  tail is well described by a log-normal distribution with a median $c=2.8$ and a scatter of $\sigma_{\log c }\approx 0.19$.
Although the distribution features an extended low-$c$ tail, we do not attempt to model this regime accurately since halos with such low concentrations are incapable of undergoing core-collapse.
Although halos with $c\gtrsim10$ are rare in the early universe, the existence of this high-$c$ tail is robustly supported by high-resolution cosmological simulations: in addition to the \VVV simulation analyzed here, the \GUREFT simulation \citep{Yung24} exhibits a very similar concentration distribution.
Therefore, for each progenitor branch in the merger trees, we draw a concentration value $c$  from this log-normal distribution.
We assume $c$ to be constant along a progenitor branch, assigning it once per branch. 
While this is a simplification, it reasonably captures the expectation that halo concentration evolves slowly. 

Third, with halo masses $\Mv(z)$ provided by the merger trees and concentrations $c$ drawn from the empirical distribution shown in \Fig{ConcentrationDistribution}, we compute the core-collapse time $\tcoll$ using \eq{CoreCollapseTime}.
The merger trees also supply the look-back time to halo formation, $\tlkbkform(z)$, for each progenitor.
For a halo with mass $\Mv(z)$ and concentration $c$, if the condition $\tcoll[\Mv(z),c]<\tlkbkform(z)$ is satisfied, we assume that the halo has undergone core-collapse and seeded a BH. 
In that case, we assign a BH seed with mass $\Mbh = 0.01\,\Mv$ to the halo and track its subsequent mass growth, as described below.
To enhance statistics, for each of the merger trees, we have 10 Monte Carlo realizations of BH seeding.

Finally, we consider BH mass growth and mergers. 
The BH growth rate is given by \citep[e.g.,][]{LoebFurlanetto13}:
\be\label{eq:BHgrowth}
\dMbhdt = \Mbh/\tE,
\ee
where the e-folding times for BH growth, $\tE$, is given by 
\be\label{eq:BHefoldingTime}
\tE = 0.45\Gyr\, \epsilon/(1-\epsilon) \lambdaE^{-1},
\ee
with $\lambdaE$ being the Eddington ratio and $\epsilon$ the radiative efficiency, typically  assumed to be 0.1 \citep{ShakuraSunyaev76}.
For simplicity, we draw $\lambdaE$ from a log-normal distribution \citep{Willott10, Xiao21}, with a median of 0.2 and a scatter of $\sigma_{\log\lambdaE}=0.3$. 
We emphasize that these choices are intended for proof-of-concept purposes, and the values are not necessarily what we advocate.

When two BH-hosting halos merge, we assume the BHs coalesce after a delay set by the dynamical-friction timescale\footnote{Here we have neglected the hardening phase of BH mergers or multi-body interactions, as well as possible effects from clumpy gas distributions at high $z$.} \citep{Boylan-Kolchin08}:
\be\label{eq:DynamicalFrictionTime}
\tmerge = 0.216 \frac{(\Mv/\mv)^{1.3}}{\ln(1+\Mv/\mv)} e^{1.9\eta} \xc \tcross
\ee 
where $\eta$ is orbital circularity, $\xc$ characterizes the orbital energy, and $\tcross = \sqrt{3/(4\pi G\Delta\rhoc)}$ is the virial crossing time.
We adopt typical values from cosmological simulations, fixing $\eta=0.5$ and $\xc=1$ \citep{Zentner05}. 
Our model tracks BH formation independently along each progenitor branch, and imposes seed inheritance or BH mergers when branches coalesce. Double-seed merger events are rare, because BH seeds form only in high-$c$ halos and for two progenitor branches to both host BHs, each must independently sample this tail. 
We neglect the possibility that a single SIDM halo can experience multiple episodes of core collapse over time due to repeated gravothermal evolution. 

Throughout this work, we have implicitly assumed that halo mass assembly histories and concentration parameters follow the same statistical distributions as in CDM. 
This assumption is supported by both physical reasoning and empirical evidence.
Physically, self-scattering primarily affects the inner halo structure, typically within a radius $r_1 \lesssim \rs$ \citep[e.g.,][]{Yang24}. 
Even during maximal core expansion, the outer halo profile remains close to the original NFW form, preserving the scale-defining parameters $\rhos$ and $\rs$. 
Empirically, cosmological SIDM simulations support this assumption. For example, \citet{Rocha13} show that both the clustering statistics and the $\Vmax$ functions of distinct halos are nearly identical between CDM and SIDM for moderate, constant cross sections. 
Consequently, it is standard practice in semi-analytical SIDM modeling to adopt CDM-based concentration-mass relations \citep[e.g.,][]{Slone23, GadNasr24, Ando25}.
That said, this assumption should be revisited in future SIDM simulations, especially those with velocity-dependent cross sections and large $\sigmaom$, where structural deviations could become more pronounced.
We also note that baryonic processes, such as adiabatic contraction \citep[e.g.,][]{Blumenthal86} or feedback-driven core formation \citep[e.g.,][]{Freundlich20}, can modify halo profiles.
However, the systems that host LRDs are expected to have low stellar content, making them less susceptible to such baryonic effects.

%-----------------------------------------------------------------

\section{The BH mass function for Little Red Dots} \label{sec:MassFunction}

Using this model, we generate random realizations of halo merger trees, populate the halos that satisfy the core-collapse condition with BH seeds, and evolve their masses over time.
We have verified that the resulting halo populations are cosmologically representative by confirming that the distribution of $\Mv(z)$ at high redshifts agrees with the theoretical HMFs at those epochs, as computed using the \hmf tool.

\begin{figure}	
\includegraphics[width=1.02\columnwidth]{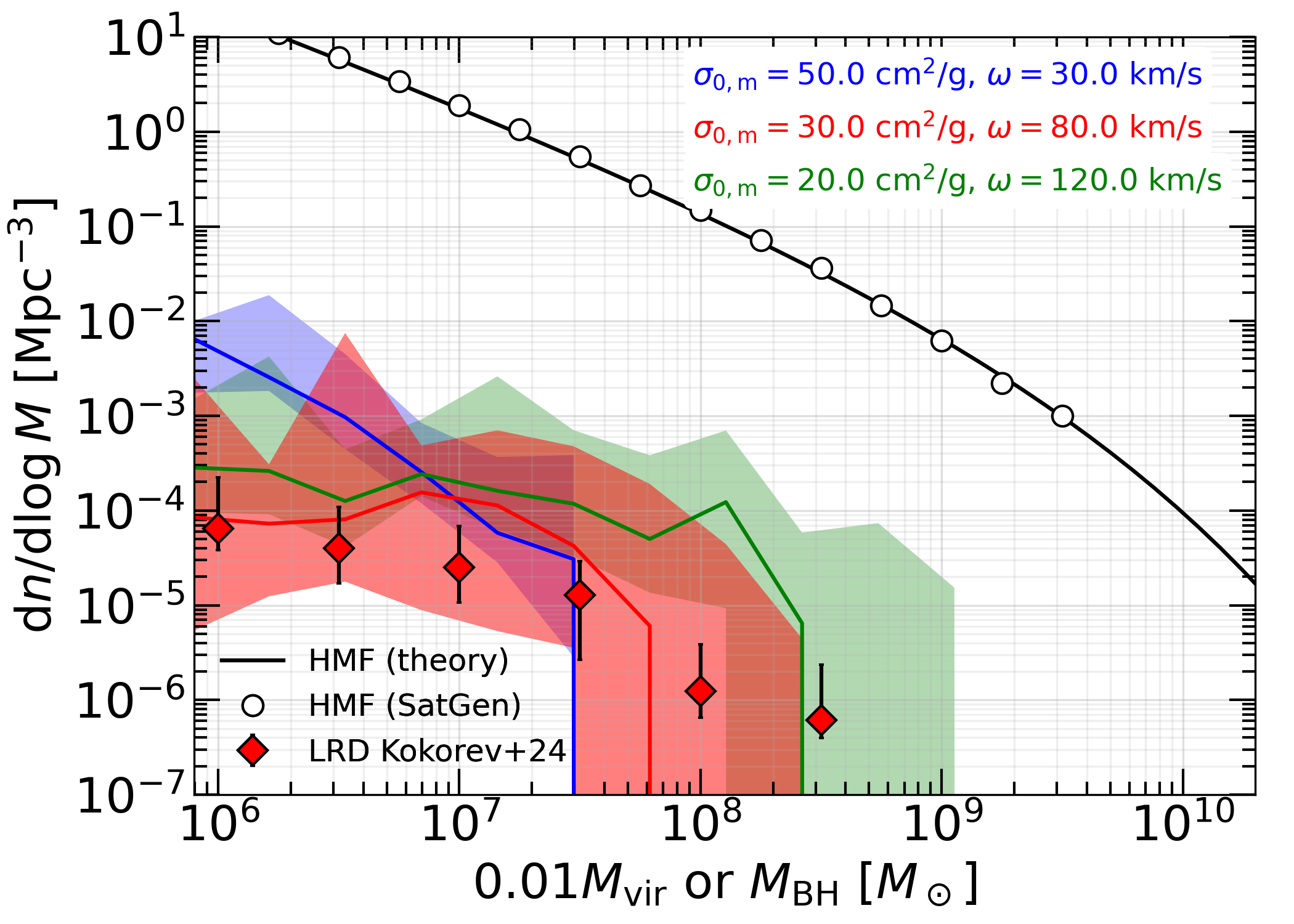}
    \caption{
    BH mass functions at $z=5$ predicted by the SIDM core-collapse and BH growth model.
    The fiducial case (red) adopts a velocity-dependent SIDM cross section with $\sigmaom=30\cm^2/\g$ and $\omega=80\kms$.
    The red solid line shows the median result across random model realizations  (see \se{model}), with the red shaded region indicating the full scatter among these realizations.
    Observationally inferred BH mass functions for LRDs at $z=4.5$-$6.5$ are shown as diamonds with error bars, adopted from \citet{Kokorev24}.
    Two alternative cross-section models (blue and green) illustrate the sensitivity of the predicted BH mass function to SIDM parameters: increasing $\sigmaom$ boosts the low-mass end; while increasing $\omega$ enhances the massive end.
     We caution against over-interpreting the cross-section values, as they are degenerate with other uncertain AGN parameters, including the Eddington ratio and duty cycle. 
    Open circles show the halo mass function (HMF) from our model realizations, in good agreement with the theoretical prediction from \hmf. }
    \label{fig:MassFunction}
\end{figure}

We find that an SIDM cross section characterized by $\sigmaom=30\cm^2/\g$ and $\omega=80\kms$, which we refer to as the fiducial model, yields BH mass functions in good agreement with those observationally inferred for LRDs.
The comparison is shown in \Fig{MassFunction}, where we adopt observational benchmarks from \citet[][see their Table 2 and Fig.6]{Kokorev24} for LRDs at $z\approx 4.5$-6.5, and compute the model predictions at a comparable redshift of $z=5$. 
We emphasize that the model is not a fit to the data, and the cross-section parameters should not be interpreted as the best-fit or preferred values, but instead just choices based on existing astrophysical constraints and lie well within the parameter space explored by recent theoretical studies (see \se{discussion}).
Several nuisance parameters that are not the focus of this study — such as the Eddington ratio and the mass fraction of a core-collapsed halo converted into a BH — can affect the BH mass function. 
However, variations in these parameters primarily shift the model predictions horizontally in \Fig{MassFunction}, rather than altering the overall shape or normalization.

The predicted BH mass function shows sensitivity to the SIDM cross-section parameters. 
To illustrate this, we compare two alternative models with ($\sigmaom$, $\omega$) =  ($50\cm^2/\g$, $30\kms$) and ($20\cm^2/\g$, $120\kms$), respectively. 
These choices are designed to highlight the impact of increasing $\sigmaom$ or $\omega$, while adjusting the other parameter to keep the overall normalization within a comparable range.
As shown in \Fig{MassFunction}, increasing $\sigmaom$ enhances the low-mass end of the BH mass function, while increasing $\omega$ produces a more extended and prominent high-mass end.
We caution against over-interpreting the apparent discrepancies between these models and the observational data.
Our fiducial model assumes a 100\% BH accretion duty cycle; more realistic, sub-unity duty cycles could easily reconcile the higher-mass or low-mass excesses seen in the alternative models.
Furthermore, the observational benchmark itself remains uncertain, with recent studies suggesting potentially lower BH mass estimates \citep[e.g.,][]{Naidu25, Graaff25}.
Given these uncertainties and degeneracies, the SIDM framework retains considerable flexibility to accommodate a range of cross sections consistent with current observations. 
\footnote{A full marginalization over nuisance parameters to constrain the SIDM cross section is beyond the scope of this study and is technically challenging. This is largely due to the computational cost of generating statistically meaningful samples of BHs, given their extremely low occupation numbers in halos. 
This rarity is already apparent from the $>4$ dex gap between the halo mass function and the BH mass function in \Fig{MassFunction}.}

%-----------------------------------------------------------------

\section{Discussion}\label{sec:discussion}

SIDM is a popular framework in near-field cosmology, especially in addressing the structures of dwarf halos \citep{SpergelSteinhardt00,Vogelsberger12,Creasey17,Roberts24} and statistics of satellite galaxies \citep{Vogelsberger19,Zeng22,ONeil23}.
Constraints on SIDM cross-section parameters are primarily derived from DM density profiles inferred through galaxy kinematic observations \citep[e.g.,][]{Slone23, YangS23, Yang24, Ando25}.
Despite recent progress, relatively broad regions of the parameter space remain viable. 
For instance, both small, nearly constant cross sections ($\sigmaom\sim 1 \cm^2/\g$, $\omega \gtrsim 100\kms$) and large cross sections at low-velocities ($\sigmaom\gtrsim 100\cm^2/\g$, $\omega\sim 10$-100$\kms$) can successfully reproduce galaxy rotation curves (RCs).
 
Particular attention has recently been drawn to the strongly interacting regime, broadly characterized by an effective cross section of $\sim20-40\cm^2/\g$ \citep{Roberts24}. 
This regime is especially attractive because such cross sections can lead to either cuspy or cored profiles, depending on the formation time and the initial halo concentration — offering a natural explanation for the observed structural diversity in dwarf galaxies. 
In fact, \citet{Turner21} and \citet{Yang23} explore even stronger interactions, using cross sections as large as $\sigmam\sim100\cm^2/\g$ at $v\sim10\kms$, and report reasonable agreement between their SIDM simulations and Milky Way satellites.

The cross section in our fiducial model -- characterized by $\sigmaom = 30 \cm^2/\g$ and $\omega = 80 \kms$ -- is not the result of fine-tuning, but rather a choice informed by existing constraints across multiple astrophysical scales. 
At bright-dwarf and Milky-Way halo scales ($\Vmax \sim 100-200\kms$), this model yields effective cross sections of $\sim 2$ and $\sim 0.3 \cm^2/\g$, respectively -- values broadly consistent with those inferred from galaxy kinematics \citep{Slone23,YangS23}. 
At cluster scales ($\Vmax \sim 500\kms$), the cross section naturally drops to $\sim 0.02 \cm^2/\g$, remaining well below the upper limits set by merging systems such as the Bullet Cluster \citep{Robertson17}. 
These parameter values fall within the viable regions defined in recent joint constraints \citep{Slone23} and are also encompassed by the broad ranges explored in SIDM modeling efforts \citep{Turner21, Nadler25}. 

While our model reproduces the observed LRD mass function for this fiducial cross section, we emphasize that this is not the unique viable choice; rather, it demonstrates that SIDM scenarios consistent with independent observational bounds can naturally give rise to early BH formation at the level required by LRD demographics.
The sensitivity of the LRD mass function to the cross-section parameters as illustrated in \Fig{MassFunction} suggests that statistics of high-$z$ BHs may provide an independent and complementary avenue for constraining SIDM cross sections — one that is orthogonal to traditional approaches based on local galaxy kinematics.

Aside from dark-matter physics and the abundance of LRDs, our model may also have implications on the clustering strength of LRDs or high-$z$ AGN. 
This is because BH seeding in our model favors early-forming, high-concentration halos, which leads to strong assembly bias. 
Interestingly, this seems to align qualitatively with preliminary observational findings on the enhanced clustering of LRDs (e.g., Zhuang et al. 2025; Tanaka et al. 2024)

%-----------------------------------------------------------------

\section{Conclusion}\label{sec:conclusion}

In this Letter, we explored the feasibility of seeding {\it massive} and {\it naked} BHs observed in LRDs through gravothermal core-collapse in SIDM halos, using a statistical, semi-analytical framework.
We characterized the properties of BH-seeding halos in terms of their concentration, mass, and formation redshift, and found that they must form early (at $z\gtrsim8.5$), be highly concentrated ($c\gtrsim10$), and reside within a specific mass range ($\Mv\sim10^{6.5-8.5}\Msun$) in order to seed BHs at the right times (during or slightly before re-ionization). 

The required high concentrations arise naturally and ubiquitously at early times in standard cosmological models, as confirmed by high-resolution simulations of early halos \citep{Wang20, Yung24}. 
The concentration distribution exhibits a median value of $c=2.8$ with a log-normal high-$c$  tail extending to $c\gtrsim10$, largely independent of halo mass and redshift. 
While the precise seeding conditions depend on the SIDM cross section, the qualitative behavior is appealing: the dark-seeding mechanism activates only at cosmic dawn, and the resulting BH seed masses naturally fall close to the expected range for LRDs, assuming a $\sim$ 1\% BH-to-halo mass ratio \citep{Feng21,Feng22,Feng25,GadNasr24}.

As a proof of concept, we showed that a modestly high SIDM cross section — $\sigmaom=30\cm^2/\g$ and $\omega=80\kms$ — yields BH mass functions in good agreement with those inferred from LRD observations. Remarkably, this parameter choice also lies within the region independently favored by SIDM constraints based on the kinematics of nearby galaxies \citep{Slone23,YangS23}.
The BH mass function exhibits sensitivity to the cross-section parameters: increasing $\sigmaom$ boosts the low-mass end, and increasing $\omega$ yields  a more extended and prominent high-mass end. 
Hence, high-$z$ BH demographics may provide an orthogonal avenue for constraining DM physics. 

Although our analysis is framed in the context of LRDs, we emphasize that the results represent a general prediction of the SIDM paradigm. Previous studies have extensively studied SIDM core-collapse since the early work of \citet{BalbergShapiro02}. 
Our contribution here takes a further step, demonstrating that within this DM framework, the formation of a cosmologically significant population of massive BHs at cosmic dawn is not only plausible — it is inevitable.

We have presented a theoretical framework to characterize this population and advocate for its inclusion in future SIDM studies. Incorporating high-$z$  BH statistics alongside local galaxy kinematics could lead to joint constraints on the particle properties of DM. Given SIDM’s compelling ability to explain the structural diversity of dwarf galaxies, we argue that it must be taken seriously as a viable, non-baryonic channel for BH seeding.

%-----------------------------------------------------------------

\begin{acknowledgments}
FJ thanks Haibo Yu, Ethan Nadler, Daneng Yang, and Joel Primack for helpful discussions. 
FJ and ZJ acknowledge support by the National Natural Science Foundation of China (NSFC, 12473007) and the Beijing Natural Science Foundation (QY23018).  
HZ is supported by the Peking University Boya Postdoctoral Fellowship.
LCH acknowledges support by the NSFC (12233001) and the National Key R\&D Program of China (2022YFF0503401).
KI acknowledges support from the NSFC (12073003, 11721303, 11991052, 12233001), the National Key R\&D Program of China (2022YFF0503401), and the China Manned Space Project (CMS-CSST-2021-A04 and CMS-CSST-2021-A06).
WXF acknowledges support from China Postdoctoral Science Foundation (2024M761594).
\end{acknowledgments}

%-----------------------------------------------------------------

\bibliography{LRD}{}
\bibliographystyle{aasjournal}

%-----------------------------------------------------------------

%\appendix



\end{document}